\documentclass{article}
\usepackage{a4wide}
\usepackage{authblk}
\usepackage{amsmath, amssymb, graphicx,psfrag,array}
\usepackage{pdfsync,color}
\usepackage[cp1251]{inputenc}
\newcommand{\triangulation}{\mathcal{T}}
\newcommand{\mcE}{\mathcal{E}}
\newcommand{\mcT}{\mathcal{T}}

\newcommand{\bfx}{{\boldsymbol x}}

\newcommand{\bi}[1]{\mbox{\boldmath $#1$}}

\def\defeq{:=}
\def\tv #1{\mbox{\boldmath $\sf #1$}}

\newcommand{\Wspace}{{\mathcal{D}\mathcal{P}}}

\def\jmp #1{[\![ #1 ]\!]}
\def\mean #1{\langle #1 \rangle}

\newcommand{\bfn}{\boldsymbol n}

\newcommand{\bfI}{\boldsymbol I}
\newcommand{\bfsig}{\boldsymbol\sigma}

\newcommand{\bfeps}{\boldsymbol\varepsilon}

\def\bbR{{\mathbb{R}}}

\def\bbR{{\mathbb{R}}}

\begin{document}
\title{Augmented Lagrangian approach to deriving discontinuous Galerkin methods for nonlinear elasticity problems}
\author[$\dagger$]{Peter Hansbo}
\author[$\ddagger$]{Mats G. Larson}
\affil[$\dagger$]{\footnotesize\it  Department of Mechanical Engineering, J\"onk\"oping University, SE-55111 J\"onk\"oping, Sweden}
\affil[$\ddagger$]{\footnotesize\it Department of Mathematics and Mathematical Statistics, Ume{\aa}~University, SE-901\,87 Ume{\aa}, Sweden}

\maketitle

\begin{abstract}
We use the augmented Lagrangian formalism to derive discontinuous Galerkin formulations for problems in nonlinear elasticity. In elasticity stress is typically a symmetric function of strain, leading to symmetric tangent stiffness matrices in Newton's method when conforming finite elements are used for discretization.  By use of the augmented Lagrangian framework, we can also obtain symmetric tangent stiffness matrices in discontinuous Galerkin methods. We suggest two different approaches and give examples from plasticity and from large deformation hyperelasticity.
\end{abstract}

\section{INTRODUCTION}

The fundamental setting for constitutive modelling in finite elasticity is that of hyperelasticity, i.e., materials for which the stress--strain relationship derives from a strain energy density function. This leads to minimisation problems from which field equations can be derived. Since the minimisation principle can then be seen as the fundamental one, it is natural to seek approximation methods also based
on minimisation, as is the case in the augmented Lagrangian method (where continuous approximations are used). In this paper, we propose a Discontinuous Galerkin (DG) method based on an augmented Lagrangian functional. The view of Nitsche's method \cite{Nit70}, underlying most DG methods, as resulting from an augmented Lagrangian has recently been explored for contact problems \cite{BuHaLa19}, obstacle problems \cite{BuHaLaSt17}, and for cut finite element methods \cite{BuHa17}. The aim of this paper is to show how the augmented Lagrangian approach can be used to derive a simple and robust Newton method for the numerical solution of nonlinear hyperelasticity problems.

There has been a great deal of interest in developing DG methods for large deformation elasticity problems, and several
forms have been suggested in the literature, cf. \cite{BaQu13,EyCeLe08a,NoRa06,EyLe06,KaLeCo15,NgPe12,NoRa06,NoRa07,OrSu07,Wh09,TrChMa15,TrChMa15b}
These are typically based on arguments different from those used in this paper, and we will return to these in Section \ref{sec:review}.

An outline of the paper is as follows. In Section \ref{sec:scalar} we consider a simple scalar model problem in order to explain the approach with as little complexity as possible; we also discuss and make comparisons with existing literature. In Section \ref{sec:finite}
we apply the approach to the case of two--dimensional antiplane shear plasticity and to three--dimensional finite elasticity, and define two variants of the linearised DG method. Finally, in Section \ref{sec:numex} we give some numerical examples for plasticity and for the Mooney--Rivlin model of large deformation elasticity.

\section{A SCALAR MODEL PROBLEM}\label{sec:scalar}

\subsection{The linear case}
A typical solid mechanics problem on a closed domain $\Omega\subset\bbR^n$, $n=2$ or 3, is to have the stress as a nonlinear function of strain. We introduce the strain vector as $\bfeps(u) = \nabla u$ for a scalar displacement field $u$, and we have the stress vector as $\bfsig = \bfsig(\bfeps(u))$. We consider hyperelasticity, when the stress can be derived from an elastic potential $\psi=\psi(\bfeps)$ such that
\begin{equation}
\bfsig = \frac{\text{d} \psi}{\text{d}\bfeps}
\end{equation}
Consider the minimisation problem
\begin{equation}\label{eq:contopt}
u = \arg\inf_{v \in H^1(\Omega)} \Psi(v)  - (f,v)_{\Omega} \mbox{ such that $u=g$ on $\partial\Omega$}
\end{equation}
where $\Psi(v)$ represents the total strain energy
\begin{equation}
\Psi(v) :=\int_\Omega\psi(\bfeps(v))\,\text{d}\Omega 
\end{equation}
The boundary condition can be enforced by use of a Lagrange multiplier $\mu$, and we then seek stationary points to the Lagrangian
\begin{equation}
\mathcal{L}(v,\mu) := \Psi(v) -
(\mu,v-g )_{\partial\Omega} 
 - (f,v)_\Omega 
 \end{equation}
The corresponding Euler equations are to seek $(u,\lambda)\in H^1(\Omega)\times H^{-1/2}(\partial\Omega)$ such that
\begin{align}
\left(\bfsig(\bfeps(u)),\bfeps(v)\right)_\Omega-(\lambda ,v)_{\partial\Omega} ={}& (f,v)\quad \forall v\in V:= H^1(\Omega)\label{eq:weak1}\\
( u,\mu)_{\partial\Omega} = {}& ( g,\mu)_{\partial\Omega} \quad \forall \mu\in Q := H^{-1/2}(\partial\Omega)
\end{align}
On strong form we derive the problem to find $u$ such that
\begin{align}
-\nabla\cdot \left(\bfsig(\nabla u)\right) ={}& f\quad \text{in $\Omega$}\\
u = {}& g\quad \text{on $\partial\Omega$}
\end{align}
We also find that 
\begin{align}
\lambda = {}& \bfn\cdot\bfsig(\nabla u) \quad \text{on $\partial\Omega$}
\end{align}
where $\bfn$ is the outward unit normal to $\partial\Omega$.

Now, Nitsche's method for the weak enforcement of boundary conditions can be derived from the following augmented Lagrangian
\begin{equation}\label{eq:ALabst}
\mathcal{L}_\text{A}(v,\mu) := \Psi(v) -
(\mu,v-g )_{\partial\Omega} + \frac{ \gamma}{2} \|v-g\|_{\partial\Omega}^2 - (f,v)_\Omega
\end{equation}
whose Euler equations are to seek $(u,\lambda)$ such that
\begin{align}
\left(\bfsig(\bfeps(u)),\bfeps(v)\right)_\Omega-(\lambda ,v)_{\partial\Omega}+\gamma (u-g,v)_{\partial\Omega} ={}& (f,v)\quad \forall v\in V\label{eq:weak1}\\
( u,\mu)_{\partial\Omega} = {}& ( g,\mu)_{\partial\Omega} \quad \forall \mu\in Q
\end{align}
where the penalty parameter $\gamma > 0$ can be chosen arbitrarily.
Consider first the case where $\bfsig = k \bfeps$ with $k$ a constant. Then Nitsche's method
is to replace $\lambda = k\bfn\cdot\nabla u$, set $\mu = k\bfn\cdot\nabla v$, and seek $u^h\in V_h$, where $V_h$ is a discrete space of piecewise continuous polynomials, such that
\begin{equation}\label{eq:Nitsche}
\left(k\nabla u^h,\nabla v\right)-(k\bfn\cdot\nabla u^h ,v)_{\partial\Omega} -( u^h, k\bfn\cdot\nabla v)_{\partial\Omega} +( h^{-1}\gamma_0u^h ,v)_{\partial\Omega}= 
 (f,v)_\Omega+ ( g,\gamma_0h^{-1}v-k\bfn\cdot\nabla v)_{\partial\Omega} 
\end{equation}
for all $v\in V_h$. Here we set $\gamma :=\gamma_0/h$, where $h$ is the local meshsize and $\gamma_0$ is a sufficiently large positive number. With these choices we have a symmetric positive definite system of equations: denoting by
\begin{equation}
A_h(u^h,v) := \left(k\nabla u^h,\nabla v\right)-(k\bfn\cdot\nabla u^h ,v)_{\partial\Omega} -( u^h, k\bfn\cdot\nabla v)_{\partial\Omega} +( h^{-1}\gamma_0u^h ,v)_{\partial\Omega}
\end{equation}
\begin{align}
\| u\|_{1/2,h,k}^2 := {}& (k h^{-1} u,u)_{\partial\Omega}\\
\| u\|_{-1/2,h,k}^2 := {}&(k h \, u,u)_{\partial\Omega}
\end{align}
we have the following inverse inequality for $v\in V_h$ 
\begin{equation}\label{inverse}
\Vert  \nabla v\Vert^2_{-1/2,h,k}\leq C_I \| \nabla v\|_{k}^2 , \quad \| \nabla v\|_{k}^2 := (k\nabla v,\nabla v)_{\Omega}
\end{equation}
We then find
\begin{align}
A_h(u,u) \geq {}& \Vert  \nabla u\Vert^2_{k}  -2 \| u\|_{1/2,h,k} \Vert \bfn\cdot\nabla u\|_{-1/2,h,k} +\gamma_0( h^{-1} u,u)_{L_2(\partial\Omega)} \\
\geq {}& \Vert  \nabla u\Vert^2_{k}  - \frac{1}{\epsilon}\Vert\bfn\cdot\nabla u \Vert_{-1/2,h,k}^2 +  ( (\gamma_0 - \epsilon k) h^{-1} u,u)_{L_2(\partial\Omega)}
\end{align}
Thus, by (\ref{inverse}), with $\epsilon > C_I$ and with the choice $\gamma_0 >  \epsilon k$ the system is positive definite.

\subsection{The non-linear case}
\subsubsection{Case of conforming methods with boundary conditions}
The question is now how to formulate a Nitsche method in the nonlinear case. The problem is that a direct
analogue to (\ref{eq:Nitsche}) loses symmetry: if we write the boundary contributions from the multipliers naively as
\begin{equation}\label{eq:Nitscheterm}
-(\bfn\cdot\bfsig(\nabla u) ,v)_{\partial\Omega} -( u, \bfn\cdot\bfsig(\nabla v))_{\partial\Omega} 
\end{equation}
then the symmetry is lost after linearisation. In the linearisation we set $u = u_n + \bar{u}$ where $\bar{u}$ is a small correction to the known field $u_n$.
We then find that $\bfsig(\bfeps(u))\approx \bfsig_n+\bfsig'_n\nabla \bar u$ where $\bfsig_n := \bfsig(\bfeps(u_n))$ and
\begin{equation}
\bfsig'_n := \left.\frac{\text{d}\bfsig}{\text{d}\bfeps}\right\vert_{u=u_n}
\end{equation}
which is typically a symmetric matrix in practical applications. Then the linearised form of (\ref{eq:Nitscheterm}) becomes
\begin{equation}
-(\bfn\cdot(\bfsig'_n\nabla \bar u) ,v)_{\partial\Omega} -( \bar u, \bfn\cdot\bfsig(\nabla v))_{\partial\Omega} -( u_n, \bfn\cdot\bfsig(\nabla v))_{\partial\Omega}-(\bfn\cdot\bfsig_n, v)_{\partial\Omega}
\end{equation}
and, crucially, the first two terms are not symmetric. There is thus no point in adding a nonlinear symmetrizing term to Nitsche's method in general.

We wish to retain symmetry in our discrete scheme in analogy with standard finite elements, and to this end we instead linearise the multiplier formulation, set $\lambda = \lambda_n +\bar{\lambda}$ with $\lambda_n$ assumed known, and we seek $(\bar{u},{\bar{\lambda}})\in V\times Q$ such that 
\begin{align}
\left(\bfsig'_n\nabla\bar{u},\nabla v\right)_\Omega-(\bar{\lambda} ,v)_{\partial\Omega} ={}& (f,v)_\Omega-\left(\bfsig_n,\nabla v\right)_\Omega+(\lambda_n ,v)_{\partial\Omega}\quad \forall v\in V\\
( \bar{u},\mu)_{\partial\Omega} = {}& ( g-u_n,\mu)_{\partial\Omega} \quad \forall \mu\in Q
\end{align}
 Now, the strong form of these equations is
\begin{align}
-\nabla\cdot \left(\bfsig'_n\nabla\bar{u}\right) ={}& f+\nabla\cdot\bfsig_n\quad \text{in $\Omega$}\\
\bar u = {}& g-u_n\quad \text{on $\partial\Omega$}
\end{align}
and 
\begin{align}\label{eq:lambda}
\bar{\lambda} = {}& \bfn\cdot\left(\bfsig'_n\nabla \bar u+\bfsig_n\right) -\lambda_n\quad \text{on $\partial\Omega$}
\end{align}
We are now free to choose $\lambda_n$ independently of $u_n$. Taking $\lambda_n = \bfn\cdot\bfsig_n$, (\ref{eq:lambda}) is simplified to 
\begin{align}\label{eq:lambdabar}
\bar{\lambda} = {}& \bfn\cdot\left(\bfsig'_n\nabla \bar u\right) \quad \text{on $\partial\Omega$}
\end{align}

Returning to the discrete augmented Lagrangian form for the linearised problem: given $(u_n^h,\lambda_n^h)\in V_h\times Q_h$, we seek $(\bar{u}^h,\bar{\lambda}^h)\in V_h\times Q_h$ such that
\begin{align}\nonumber
\left(\bfsig'_n\nabla\bar{u}^h,\nabla v\right)_\Omega-(\bar{\lambda}^h ,v)_{\partial\Omega}+(\gamma_0h^{-1} \bar{u}^h,v)_{\partial\Omega} ={}& (f,v)_\Omega  -\left(\bfsig_n,\nabla v\right)_\Omega+(\lambda_n ,v)_{\partial\Omega}\\  &-(\gamma_0h^{-1} (u_n-g),v)_{\partial\Omega} \quad \forall v\in V_h\\
( \bar{u}^h,\mu)_{\partial\Omega} = {}& ( g-u_n,\mu)_{\partial\Omega} \quad \forall \mu\in Q_h
\end{align}
where we use the notation $\bfsig'_n = \bfsig'(u^h_n)$, $\bfsig_n=\bfsig(u^h_n)$. Choosing now $\bar{\lambda}^h = \bfn\cdot\left(\bfsig'_n\nabla \bar{u}^h\right)$ and $\mu = \bfn\cdot\left(\bfsig'_n\nabla v\right)$ we obtain the following Nitsche--Newton method: given $u_0^h$, set $n=0$ and seek $\bar{u}^h\in V_h$ such that
\begin{align}\nonumber
\left(\bfsig'_n\nabla\bar{u}^h,\nabla v\right)_\Omega-(\bfn\cdot\bfsig'_n\nabla \bar{u}^h ,v)_{\partial\Omega} -( \bar{u}^h, \bfn\cdot\bfsig'_n\nabla v)_{\partial\Omega} +( h^{-1}\gamma_0\bar{u}^h ,v)_{\partial\Omega}= \\
 (f,v)_\Omega-\left(\bfsig_n,\nabla v\right)_\Omega +(\bfn\cdot\bfsig_n ,v)_{\partial\Omega} + ( g-u_n^h,h^{-1}\gamma_0v-\bfn\cdot\bfsig'_n\nabla v)_{\partial\Omega} \label{eq:nitchenewton}
\end{align}
for all $v\in V_h$, set $u^h_{n+1}=u^{h}_{n}+\bar{u}^h$, $n\mapsto n+1$ and repeat until convergence. This results in a sequence of symmetric discrete problem if $\bfsig'_n$ is symmetric. 

It is instructive to look at the corresponding scheme in the linear case when
$\bfsig'_n = k\bi I$ and $\bfsig_n = k\nabla u^h_n$: find $u^h_{n+1}\in V_h$ such that
\begin{align}\nonumber
\left(k\nabla u^h_{n+1},\nabla v\right)_\Omega-(k\bfn\cdot\nabla u^h_{n+1} ,v)_{\partial\Omega} -( u^h_{n+1}, k\bfn\cdot\nabla v)_{\partial\Omega} +( h^{-1}\gamma_0u^h_{n+1} ,v)_{\partial\Omega}= \\
 (f,v)_\Omega  + ( g,h^{-1}\gamma_0v-k\bfn\cdot\nabla v)_{\partial\Omega} 
\end{align}
for all $v\in V_h$, which coincides with (\ref{eq:Nitsche}).
In analogy with the linear case, we write
\[
A_h^n(u,v) := \left(\bfsig'_n\nabla u,\nabla v\right)_\Omega-(\bfn\cdot\bfsig'_n\nabla u ,v)_{\partial\Omega} -( u, \bfn\cdot\bfsig'_n\nabla v)_{\partial\Omega} +( h^{-1}\gamma_0u ,v)_{\partial\Omega}
\]
and for coercivity of the linearised Nitsche problem we clearly need $\bfsig'_n$ to be positive definite. 
We further need to determine $C_I$ from the inverse inequality
\begin{equation}\label{inverse2}
(h \bfsig'_n\nabla v,\nabla v)_{\partial\Omega}\leq C_I (\bfsig'_n\nabla v,\nabla v)_{\Omega} 
\end{equation}
Setting
\begin{align}
\| u\|_{1/2,h,n}^2 := {}& (h^{-1} \bfn\cdot\bfsig'_n\cdot\bfn\, u,u)_{\partial\Omega}\\
\| \nabla u\|_{-1/2,h,n}^2 := {}&( h \bfsig'_n \nabla u,\nabla u)_{\partial\Omega}
\end{align}

\begin{align}
A_h^n(u,u) \geq {}& \left(\bfsig'_n\nabla u,\nabla u\right)_\Omega  -2 \| u\|_{1/2,h,n} \Vert \nabla u\|_{-1/2,h,n} +\gamma_0( h^{-1} u,u)_{\partial\Omega} \\
\geq {}&  \left(\bfsig'_n\nabla u,\nabla u\right)_\Omega  - \frac{1}{\epsilon}\Vert \nabla u \Vert_{-1/2,h,n}^2 +  ( (\gamma_0 - \epsilon\bfn\cdot\bfsig'_n\cdot\bfn) h^{-1} u,u)_{\partial\Omega}
\end{align}
so that with the choice $\gamma_0 >  C_I\bfn\cdot\bfsig'_n\cdot\bfn$ or, without explicit dependence on $\bfn$, 
$\gamma_0 >  C_I \sigma_n$, where $\sigma_n$ is the largest eigenvalue of $\bfsig'_n$,  the form $A_h^n(u,u)$ is coercive.
It should be noted that the inequality (\ref{inverse2}) in fact only has to hold on elements $T$ adjacent to the boundary, so that
\[
(h \bfsig'_n\nabla v,\nabla v)_{T\cap\partial\Omega}\leq C_{I,T} (\bfsig'_n\nabla v,\nabla v)_{T} 
\]
giving different constants for different elements. 
\subsubsection{Case of discontinuous Galerkin methods, classical approach}

We now wish to find a related discrete minimisation problem for discontinuous approximations. For ease of presentation we assume here that $u=0$ on $\partial\Omega$, the nonzero case is handled as above.
We introduce a a subdivision $ \triangulation =\{ T\}$ of $\Omega$ of 
$\Omega$ into a geometrically conforming finite element mesh.  
For simplicity, we assume 
that $\triangulation$ is quasiuniform. Let
\[\begin{array}{l}
\Wspace = \{c\in [L^2(\Omega)]:~\text{$v \vert_T \in P^q(T)$ for all 
$T \in \triangulation$} \},
\end{array}\]
be the space of piecewise discontinuous polynomials of degree $q$. The set 
of faces (edges) in the mesh is denoted by $\mcE =\{ E \}$ and we split $\mcE$ 
into two disjoint subsets 
\begin{equation*}
  \mcE = \mcE_I \cup \mcE_D ,
\end{equation*}
where $\mcE_I$ is the set of faces in the interior of $\Omega$, 
$\mcE_D$ is the set of faces on the boundary 
$\partial \Omega$. Further, with each edge we 
associate a fixed unit normal $\bi n$  such that for edges on the 
boundary $\bi n$ is the exterior unit normal. We denote the jump of 
a function $v \in \Wspace$ at a face $E$ by 
$ \jmp{ v} =  v^+- v^-$ for $E \in \mcE_I$ and 
$ \jmp{ v} =  v^+$ for $E \in \mcE_D$, and the average 
$\left<   v\right> = ( v^+ +  v^-)/2$ for $E \in \mcE_I$ and 
$\left<   v\right> =  v^+$ for $E \in \mcE_D$, where 
$ v^{\pm} = \lim_{\epsilon\downarrow 0}  v(\bfx\mp \epsilon\,\bi n)$ 
with $\bfx\in E$. 
With $|E|$ the measure of $E$, we define the meshsize $h$ by 
\begin{equation}
h|_E = \min
 \left( |T^+| /|E|, |T^-| /|E|\right)  \quad \text{for $E = \partial T^+ \cap
\partial T^-$},
\end{equation}
with $|T|$ the measure of $T$, on each face. 

The augmented Lagrangian method is then to seek $u$ and $\lambda$ that give stationarity of the functional
\begin{equation}
\mathcal{L}_\text{A} (v,\mu)\defeq \Psi( v)-\sum_{E\in\mcE} (\mu,\jmp{v})_E +\frac{\gamma}{2}\sum_{E\in\mcE}( \jmp{v},\jmp{ v})_E-(f, u)_\Omega
\label{eq:DGAug}
\end{equation}
For the internal faces, we now need to define average tractions and linearised tractions. We can then choose between $\mean{\bfsig_n}$, $\mean{\bfsig'_n}$, and
$\tilde{\bfsig}_n := \bfsig(\mean{u_n})$, $\tilde{\bfsig}'_n := \bfsig'(\mean{u_n})$, both are consistent. In cases where the stresses are smooth, the choice of traction is not crucial and
we then choose to use $\tilde{\bfsig}_n$, $\tilde{\bfsig}'_n$ as this involves fewer computations.
Proceeding as above, defining
\[
A_h^n(u,v) := \sum_{T\in\mcT}\left(\bfsig'_n\nabla u,\nabla v\right)_T-\sum_{E\in\mcE}(\bfn\cdot\tilde{\bfsig}'_n\mean{\nabla u} ,\jmp{v})_{E} -\sum_{E\in\mcE}( \jmp{u}, \bfn\cdot\tilde{\bfsig}'_n\mean{\nabla v})_{E} +\sum_{E\in\mcE}( h^{-1}\gamma_0\jmp{u} ,\jmp{v})_{E}
\]
 we obtain the following Nitsche--Newton method: given $u_0^h$, set $n=0$ and seek $\bar{u}^h\in \Wspace$ such that
\begin{align}\nonumber
A_h^n(\bar{u}^h, v)= {}&
 (f,v)_\Omega-\sum_{T\in\mcT}\left(\bfsig_n,\nabla v\right)_T +\sum_{E\in\mcE}(\bfn\cdot\tilde{\bfsig}_n,\jmp{v})_{E}   \\ & -\sum_{E\in\mcE}( \jmp{u_n^h},h^{-1}\gamma_0\jmp{v}-\bfn\cdot\tilde{\bfsig}'_n\mean{\nabla v})_{E} \label{eq:nitchenewton2}
\end{align}
for all $v\in \Wspace$, set $u^h_{n+1}=u^{h}_{n}+\bar{u}^h$, $n\mapsto n+1$ and repeat until convergence.

\subsubsection{Case of discontinuous Galerkin methods, hybridized approach}
In the case of non-smooth stresses, as for example in plasticity problems, our experience is, however, that convergence problems can occur
with both types of averaging of the tractions. For such problems, we therefore propose to use an alternative implementation of the DG method which avoids averaging of tractions. We then do not consider sums over edges; with $\bfn_T$ the outward pointing normal on $T$ and $h\vert_{\partial T\cap E} = \vert T\vert/\vert E\vert$, we instead write, in the linear case: find ${u}^h\in \Wspace$ such that
\begin{align}\nonumber
(f,v)_{\Omega} = &{} \sum_{T\in\mcT}\Bigl((k\nabla u^h,\nabla v)_T-(k \bfn_T\cdot\nabla u^h ,{v}-\mean{v})_{\partial T} -( {u^h}-\mean{u^h}, k\bfn_T\cdot{\nabla v})_{\partial T} \\
& +( h^{-1}\gamma_0({u^h}-\mean{u^h}) ,{v}-\mean{v})_{\partial T}\Bigr) \quad\forall v\in \Wspace
\end{align}
This formulation underpins hybrid DG methods \cite{BuElHaLaLa19}, where $\mean{u^h}$ is replaced by an independent ``hybrid'' variable which is particularly useful for model coupling purposes.\cite{BuHaLaLa19,BuHaLa20} The derivation from an augmented Lagrangian method follows along the same lines as the previous formulation, and we can write the Newton iteration scheme as that of finding $\bar{u}^h\in \Wspace$ such that
\begin{align}\nonumber
\hat{A}_h^n(\bar{u}^h, v)= {}&
 (f,v)_\Omega  -\sum_{T\in\mcT}\left(\left(\bfsig_n,\nabla v\right)_T +(\bfn_T\cdot{\bfsig}_n,{v}-\mean{v})_{\partial T} \right)  \\ 
 & - \sum_{T\in\mcT}( {u_n^h}-\mean{u_n^h},h^{-1}\gamma_0({v}-\mean{v})-\bfn_T\cdot{\bfsig}'_n\nabla v)_{\partial T}\label{eq:nitchenewton3}
\end{align}
for all $v\in \Wspace$, where now 
\begin{align*}
\hat{A}_h^n(u,v) := {}& \sum_{T\in\mcT}\Bigl(\left(\bfsig'_n\nabla u,\nabla v\right)_T-(\bfn_T\cdot{\bfsig}'_n\nabla u ,{v}-\mean{v})_{\partial T} -( {u}-\mean{u}, \bfn_T\cdot{\bfsig}'_n\nabla v)_{\partial T} \\
& +( h^{-1}\gamma_0({u}-\mean{u} ),{v}-\mean{v})_{\partial T}\Bigr)
\end{align*}

\subsection{A brief review of previous work}\label{sec:review}

Eyck and Lew, \cite{EyLe06} in what appears to be the first paper on DG for large deformations, introduce discontinuous approximations of derivatives on the discrete level already in the elastic potential, which results in a symmetric linearized problem but requires lifting operators (from edges of elements to element interiors) to be defined and used at additional computational cost. The same approach is used in their subsequent related work,\cite{EyCeLe08a,EyCeLe08b} and in the hybridized DG of Kabaria, Lew, and Cockburn.\cite{KaLeCo15}

Ortner and S\"uli\cite{OrSu07} suggest
an {\it incomplete}\/ discontinuous Galerkin method (without the addition of a symmetrizing term),
leaving the resulting scheme unsymmetric even in the linear case. This approach was also used by Dolej\v{s}\'{\i}\cite{Do08} for nonlinear diffusion and by Liu, Wheeler, and Yotov\cite{LiWhYo13} for finite elastoplasticity. It is also used in a hybrid DG setting by Terrana et al.\cite{Ter19} and Wulfinghoff et al.\cite{WuBa17}

Noels and Radovitzky\cite{NoRa06} apply the Hu--Washizu principle with independent fields for deformation, stress and strain. They then show how to introduce further approximations so as to eliminate the auxiliary stress and strain fields. The resulting linearized scheme is unsymmetric. This approach was later extended to dynamics with explicit time-stepping.\cite{NoRa08}

Whiteley\cite{Wh09} uses the symmetric form in the nonlinear setting, which, as mentioned above, still leads to a nonsymmetric linearized scheme.

Baroli and Quarteroni\cite{BaQu13} use a Lagrange multiplier technique intended for incompressible hyperelasticity where mixed methods are natural. The linearized system is unsymmetric, apart from the pressure--dependent terms.

Closest to our approach is the the Variational Multiscale Discontinuous Galerkin method (VMDG) of Truster, Chen, and Masud, \cite{TrChMa15,TrChMa15b} who also start with a Lagrange multiplier approach.
This is then stabilized by a variational multiscale method which allows for the elimination of the multiplier and automatically gives stabilizing terms similar to ours. An important difference is that the multiscale method gives rise to an additional term involving third derivatives of
the strain energy (later suggested to be dropped\cite{TrChMa15b}) not present in our approach. We argue that the augmented Lagrangian route
presented herein is more direct and closer to the original Nitsche concept.

\section{APPLICATIONS}\label{sec:finite}

\subsection{Antiplane shear plasticity}

As a simple scalar model, we consider the antiplane shear problem of finding the displacement $u$ and stress $\bi{\sigma}$ such
that $-\nabla\cdot\bi{\sigma} = f$ in $\Omega\subset \bbR^2$, with $u=0$ on $\partial\Omega$, and
\[
\bi{\sigma} = \left\{\begin{array}{c}
G\nabla u \quad\text{if}\quad \vert G\nabla u \vert \leq \sigma_Y,\\[4mm]
\displaystyle\frac{\sigma_Y}{\vert\nabla u\vert}\nabla u\quad\text{if}\quad \vert G\nabla u \vert >
\sigma_Y,\end{array}\right.
\]
where $\sigma_Y$ is a constitutive parameter (the yield stress) and $G$ is the shear modulus. Thus, in the elastic range $\bi{\sigma}'=G\bfI$ and in the plastic range 
\[
\bi{\sigma}' =\frac{\sigma_Y}{\vert \nabla u\vert^{3}}\begin{bmatrix}
\displaystyle\left(\frac{\partial u}{\partial y}\right)^2 & \displaystyle -\frac{\partial
u}{\partial y}\frac{\partial u}{\partial x}\cr \displaystyle -\frac{\partial u}{\partial
y}\frac{\partial u}{\partial x} &\displaystyle \left(\frac{\partial u}{\partial x}\right)^2\end{bmatrix} 
\] 
Typically, one also takes into account the fact that the linearization is valid only at the yield surface and set
\begin{equation}\label{tangentstiff}
\bi{\sigma}' :=  \frac{G}{\vert \nabla
u\vert^{2}}\begin{bmatrix}
\displaystyle\left(\frac{\partial u}{\partial y}\right)^2 & \displaystyle -\frac{\partial
u}{\partial y}\frac{\partial u}{\partial x}\cr \displaystyle -\frac{\partial u}{\partial
y}\frac{\partial u}{\partial x} &\displaystyle \left(\frac{\partial u}{\partial x}\right)^2\end{bmatrix} 
\end{equation}
For this problem, the smoothness of the displacement field 
is only such that it has bounded variation,\cite{Str79} and we cannot expect a smooth stress field; the use of average stresses in 
the DG formulation can then lead to elements flipping in and out of the plastic zone, leading to convergence problems. Thus we advocate the form (\ref{eq:nitchenewton3}), which in our experience does not suffer from this effect.

We remark that the formulation as a 

\subsection{Finite elasticity}
\subsubsection{Problem formulation}
We also consider the more general problem of nonlinear elasticity, analogous to (\ref{eq:contopt}), seeking the displacement field $\bi u(\bi X)$ on the reference (undeformed) domain $\Omega\ni\bi X$ such that the potential
\begin{equation}
    \Pi(\bi u)\defeq\Psi(\bi u)-l(\bi u)
\label{eqPi}
\end{equation}
Consider the minimisation problem
\begin{equation}\label{eq:contopt}
\bi u = \arg\inf_{\bi v \in [H^1(\Omega)]^3} \Psi(\bi v)  - (\bi f, \bi v)_{\Omega} \mbox{ such that $\bi v=\bi 0$ on $\partial\Omega$}
\end{equation}
where $F(\bi v)$ represents the total strain energy
\begin{equation}
\Psi(\bi v) :=\int_\Omega\psi(\bi F(\bi v))\,\text{d}\Omega 
\end{equation}
becomes stationary.
Here $\psi$ is the reference volume specific free elastic energy, expressed pointwise in terms of the deformation gradient
\begin{equation}
    \bi F \defeq \bi I + \bi H, \quad \bi H \defeq \bi v \otimes \bi \nabla 
\end{equation}
where $\bi I$ denotes the second order identity tensor and $\bi H$ is the displacement gradient.
A hyper-elastic model is now defined by formulating the first Piola Kirchhoff stress tensor $\bi P=\bi P(\bi F)$ as
\begin{equation}
    \bi P \defeq \frac{{\rm d}\psi}{{\rm d}\bi F}
\end{equation}
We may thus formulate the problem as that of finding $\bi u\in W:=\{\bi v\in[H^1(\Omega)]^d, \, \bi v={\bi 0} \,\, \text{on } \partial\Omega \}$ such that
\begin{equation}
    a(\bi u, \bi v)=l( \bi v) \quad \forall \bi v\in W,
    \label{eqweak}
\end{equation}
where 
\begin{equation}
    a(\bi u,\bi v)= \int_\Omega \bi P:\left[ \bi v\otimes \bi \nabla \right] {\rm d}\Omega.
\end{equation}
and
\begin{equation}
    l(\bi v)= \int_\Omega {\bi f} \cdot \bi v \,{\rm d}\Omega 
\end{equation}

%
\subsubsection{Non-linear discrete solution to the weak problem}
For the solution of (\ref{eqweak}), we return to the Newton framework, and here we assume a smooth stress field allowing for the use of a formulation related to (\ref{eq:nitchenewton2}). We introduce an iteration $\bi u_n\in W$ and define the weak residual as
\begin{eqnarray}
    r(\bi v,\bi w)\defeq l(\bi w)-a(\bi v,\bi w).
\end{eqnarray}
Newton updates $\bi u_{n+1}-\bi u_{n} =: \bar{\bi u}\in W$ can now be solved for such that
\begin{eqnarray}
    ({\tv L}:\left(\bar{\bi u}\otimes \bi \nabla \right),\bi v\otimes \bi \nabla)_{\Omega} = r(\bi u_{n}, \bi v) , \quad \forall \bi v \in W
\end{eqnarray}
where we introduced the tangent stiffness tensor
\begin{equation}
    {\tv L}\defeq \frac{{\rm d}\bi P}{{\rm d}\bi F}.
\label{eqsecond}
\end{equation}

The augmented Lagrangian method in this case is to seek $\bi u$ and $\bi\lambda$ that give stationarity of the functional
\begin{equation}
    \Pi_h(\bi u)\defeq\Psi(\bi u)-\sum_{E\in\mcE} (\bi\lambda,\jmp{\bi u})_E +\frac{1}{2}\sum_{E\in\mcE}( \bi S\jmp{\bi u},\jmp{\bi u})_E-l(\bi u)
\label{eqPiAug}
\end{equation}
where $\bi\lambda$ is a vector-valued multiplier and $\bi S$ is a positive definite symmetric penalty matrix. Taking the first variation yields
\begin{equation}
    a(\bi u, \bi v;\bi\lambda,\bi\mu)= \sum_{T\in\mcT}(\bi P,\left( \bi v\otimes \bi \nabla \right))_T -\sum_{E\in\mcE}(\bi\mu,\jmp{\bi u})_E-\sum_{E\in\mcE}(\bi\lambda,\jmp{\bi v})_E+\sum_{E\in\mcE}(\bi S\jmp{\bi u},\jmp{\bi v})_E
\end{equation}
and the linearised form is to find $(\bar{\bi u},\bar{\bi\lambda})$, given $(\bi u_n,\bi\lambda_n)$, such that
\begin{align}\nonumber
 \sum_{T\in\mcT}\int_{T} \left( \bi v\otimes \bi \nabla \right):{\tv L}:\left(\bar{\bi u}\otimes \bi \nabla \right) {\rm d}x-\sum_{E\in\mcE}(\bi\mu,\jmp{\bar{\bi u}})_E-\sum_{E\in\mcE}(\bar{\bi\lambda},\jmp{\bi v})_E+\sum_{E\in\mcE}(\bi S\jmp{\bar{\bi u}},\jmp{\bi v})_E = \\
 l(\bi v) -a_h(\bi u_n, \bi v;\bi\lambda_n,\bi\mu) 
\end{align}
Identifying $\bi\lambda_n\vert_E$ with the traction on the edge,
\begin{equation}
\bi\lambda_n\vert_E = \bi P_n\vert_E \cdot \bi N
\end{equation}
and approximating
\begin{equation}
\bi P_n\vert_E \approx \bi P(\langle\bi F_n\rangle)
\end{equation}
we have, following (\ref{eq:lambdabar}),
\begin{equation}
\bar{\bi\lambda}\cdot\jmp{\bi v} = \left(\jmp{\bi v}\otimes \bi N\right): {\tv L}(\mean{\bi F_n}):\mean{\bar{\bi u}\otimes \bi \nabla} 
\end{equation}
Choosing now $\bi\mu$ so that
\begin{equation}
\bi\mu\cdot\jmp{\bar{\bi u}} = \left(\jmp{\bar{\bi u}}\otimes \bi N\right): {\tv L}(\mean{\bi F_n}):\mean{\bi v\otimes \bi \nabla} 
\end{equation}
and $\bi S$ such that
\begin{equation}
(\bi S\jmp{\bar{\bi u}})\cdot\jmp{\bi v} = \frac{\gamma_0}{h} \left(\jmp{\bi v}\otimes \bi N\right): {\tv L}(\mean{\bi F_n}):\left(\jmp{\bar{\bi u}}\otimes \bi N\right)
\end{equation}
we obtain the symmetric linearised form (coercive on $[\Wspace]^d$ if $\gamma_0$ is chosen sufficiently large) 
\begin{align}
  A_h^n(\bar{\bi u},\bi v)  :={}& \sum_{T\in\mcT}\int_{T} \left( \bi v\otimes \bi \nabla \right):{\tv L}:\left(\bar{\bi u}\otimes \bi \nabla \right){\rm d}x \nonumber \\
    & - \sum_{E\in\mcE}\int_E \left(\jmp{\bi v} \cdot \left({\tv L}(\mean{\bi F}):\mean{\bar{\bi u}\otimes \bi \nabla}\right)\cdot\bi N
    +\jmp{\bar{\bi u}} \cdot \left({\tv L}(\mean{\bi F}):\mean{ \bi v\otimes \bi \nabla}\right)\cdot\bi N \right){\rm d}s  \nonumber \\
    & +\sum_{E\in\mcE}\int_E \frac{\gamma_0}{h} \left(\jmp{\bi v}\otimes \bi N\right): {\tv L}(\mean{\bi F}):\left(\jmp{\bar{\bi u}}\otimes \bi N\right) {\rm d}s.
\end{align}
and the Nitsche--Newton iterations analogous to (\ref{eq:nitchenewton2}) are to find the Newton update $\bar{\bi u}^h\in[\Wspace]^d$, given $\bi u_n^h\in [\Wspace]^d$, such that
\begin{align}\nonumber
A_h^n(\bar{\bi u}^h,\bi v)={}& l(\bi v)-\sum_{T\in\mcT}\int_{T}\bi P(\bi F_n):\left( \bi v\otimes \bi \nabla \right) {\rm d}x+ \sum_{E\in\mcE}\int_E \bi P(\langle\bi F_n\rangle):\left(\jmp{{\bi v}}\otimes \bi N\right)\text{d}s \\ \nonumber
& -\sum_{E\in\mcE}\int_E \frac{\gamma_0}{h} \left(\jmp{{\bi u}^h_n}\otimes \bi N\right): {\tv L}(\mean{\bi F}):\left(\jmp{\bi v}\otimes \bi N\right) {\rm d}s \\
& +\sum_{E\in\mcE}\int_E\jmp{{\bi u}^h_n} \cdot \left({\tv L}(\mean{\bi F}):\mean{\bi v\otimes \bi \nabla}\right)\cdot\bi N \,{\rm d}s
\end{align}
for all $\bi v\in [\Wspace]^d$.
\section{NUMERICAL EXAMPLES}\label{sec:numex}

\subsection{Antiplane shear plasticity}

We take an example from Johnson and Hansbo,\cite{JoHa92} with $\Omega = (0,1)\times(0,1)$, $\sigma_Y=1$, $G=1$, and 
\[
f(\bfx) = \frac{15}{2}\sin{\pi x_1}\sin{\pi x_2}
\]
with boundary conditions $u=0$ on $\partial\Omega$. We use $\gamma_0 = 10^2G$ as stabilization parameter.

\begin{figure}
	\begin{center}
	\includegraphics[scale=0.3]{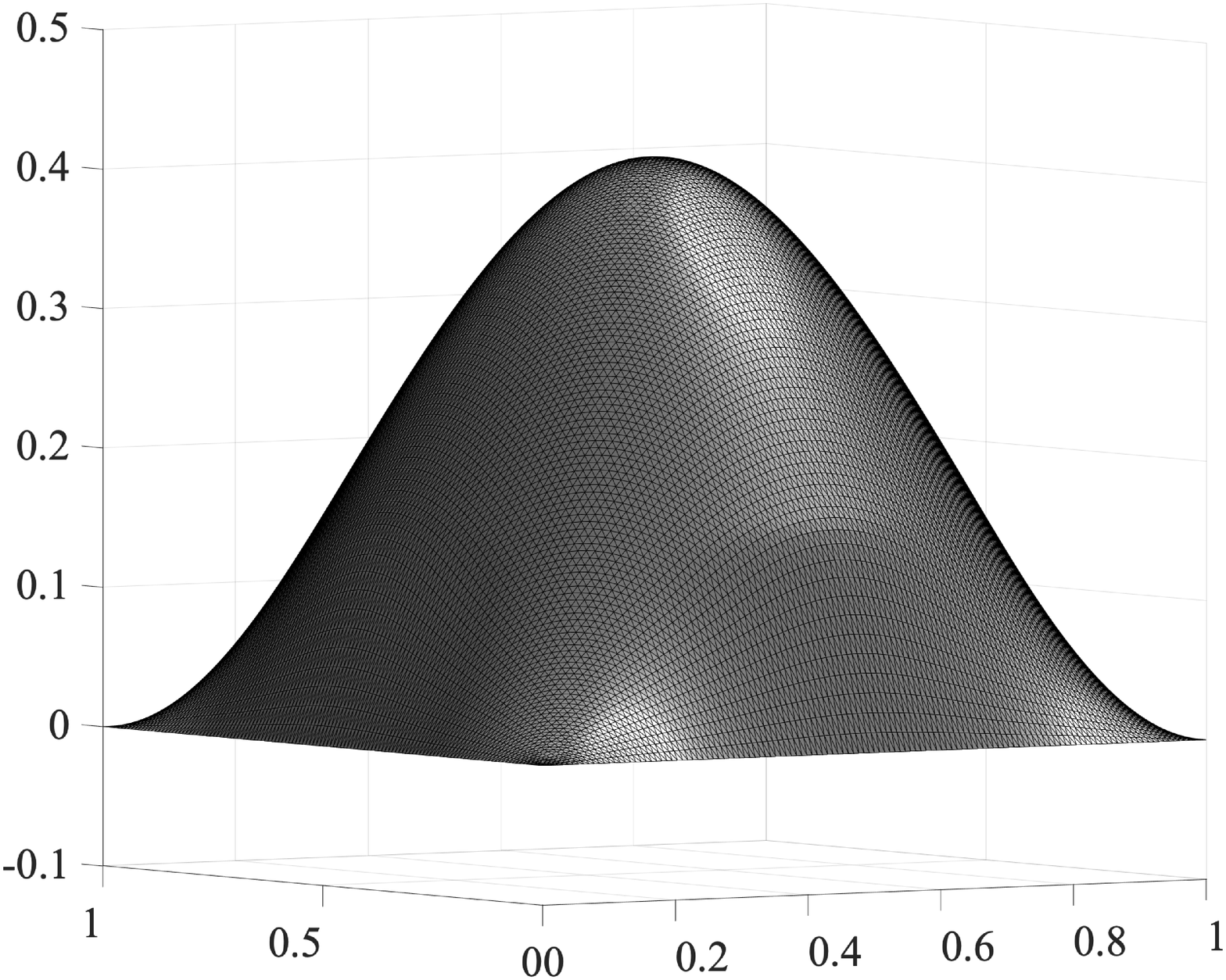}
	\end{center}
\caption{Displacement field for the plasticity problem.\label{plast:disp}}
\end{figure}

\begin{figure}
	\begin{center}
	\includegraphics[scale=0.2]{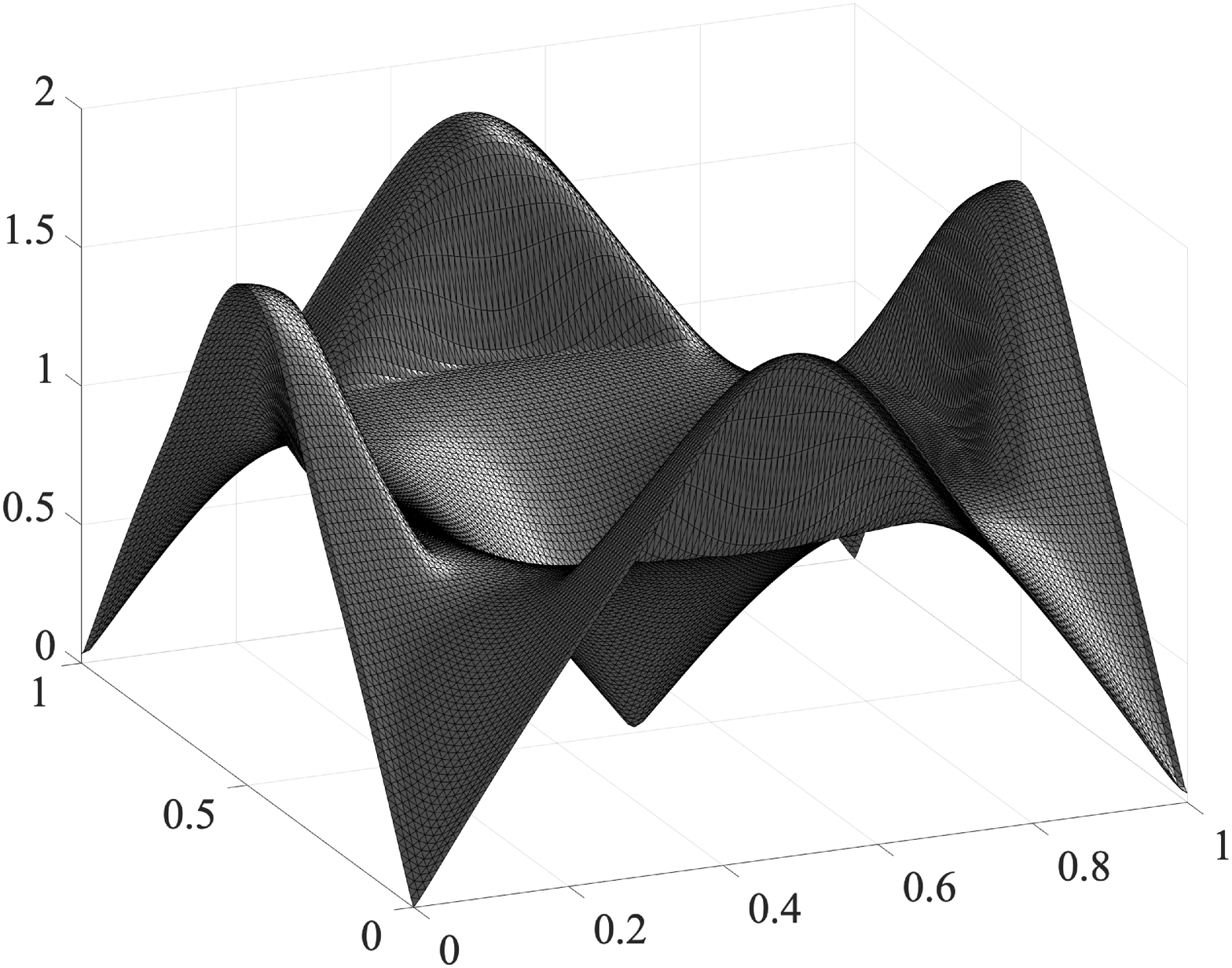}\includegraphics[scale=0.2]{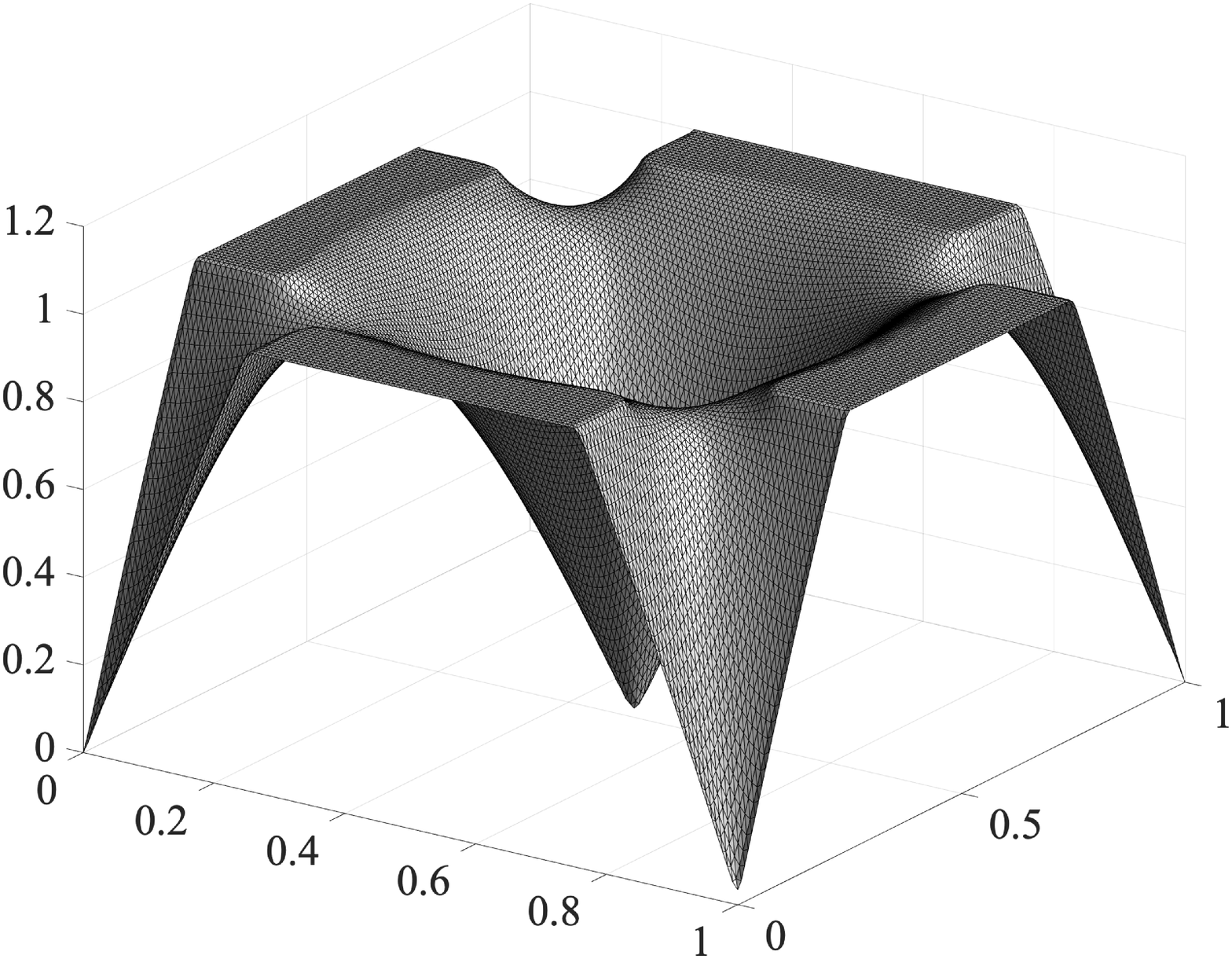}
	\end{center}
\caption{Gradient field (left) and stress (right) for the plasticity problem.\label{plast:stress}}
\end{figure}

In Fig \ref{plast:disp} we show the computed displacements, in Fig. \ref{plast:stress} we show the norm of the gradient of displacement and the stresses (projected onto the space of continuous $P^1$ elements). We note that the displacement $u$ is continuous with $\vert\nabla u\vert$ bounded.

\subsection{Large deformation elasticity}

In our numerical examples, we use an isotropic Mooney--Rivlin model in which we choose Young's modulus $E$ and Poisson's ratio $\nu$, and define
$K:=E/(3(1-2\nu))$, $\mu :=E/(2(1+\nu))$, and $\mu_1= \mu_2 = \mu/2$. Then the Mooney--Rivlin strain energy density is given by
\begin{equation}
\psi({\bi F}):=\frac12 \mu_1 J^{-2/3}I_1 +\frac12 \mu_2 J^{-4/3}I_2 +\frac12 K(J-1)^2
\end{equation} 
where $J=\text{det $\bi F$}$ and $I_1, I_2$ are the first and second invariants of the left Cauchy--Green deformation tensor $\bi b := \bi F \bi F^{\rm T}$.

In all examples we used $E=200$ GPa, $\nu = 0.33$, and $\gamma_0=10^2$. We compare a piecewise linear DG solution to a standard conforming $P^1$ finite element method on the same mesh for three different typical modes of deformation: twisting, stretching, and bending. The solutions are close in all examples.

\subsubsection{Twisting}

\begin{figure}
	\begin{center}
	\includegraphics[scale=0.3]{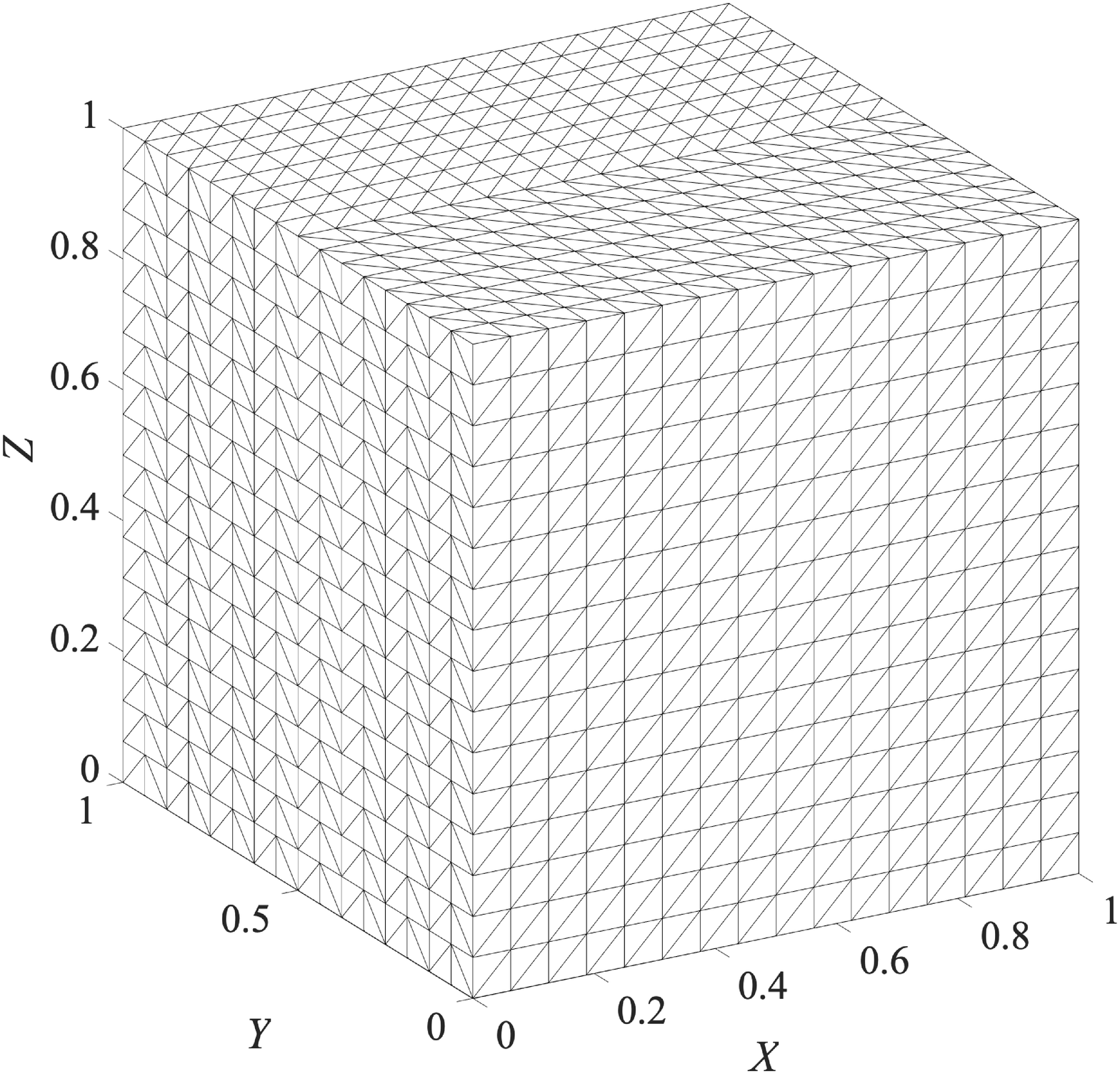}
	\end{center}
\caption{Mesh used for twisting computations.\label{fig:twistmesh}}
\end{figure}

\begin{figure}
	\begin{center}
	\includegraphics[scale=0.2]{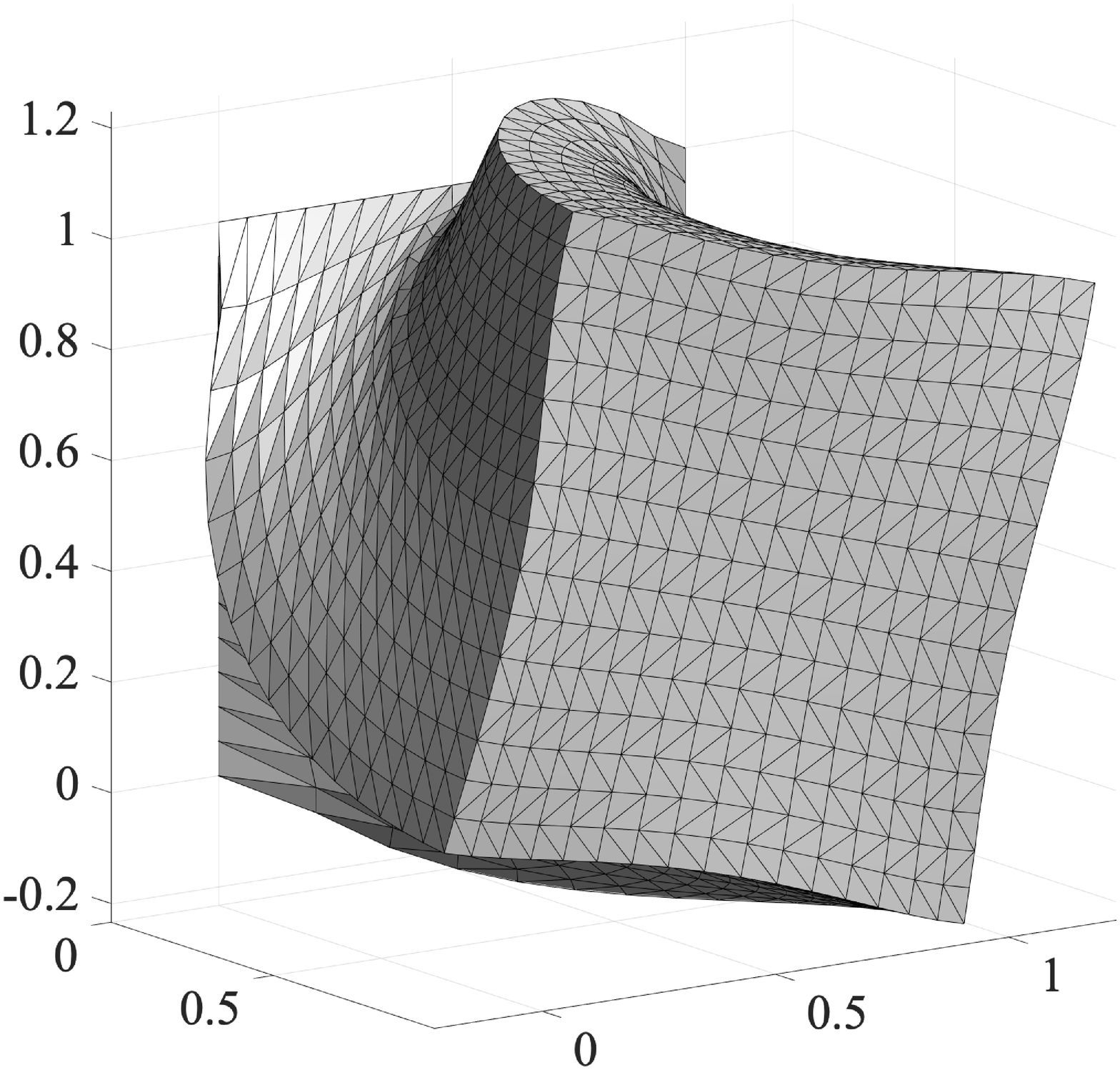}\includegraphics[scale=0.2]{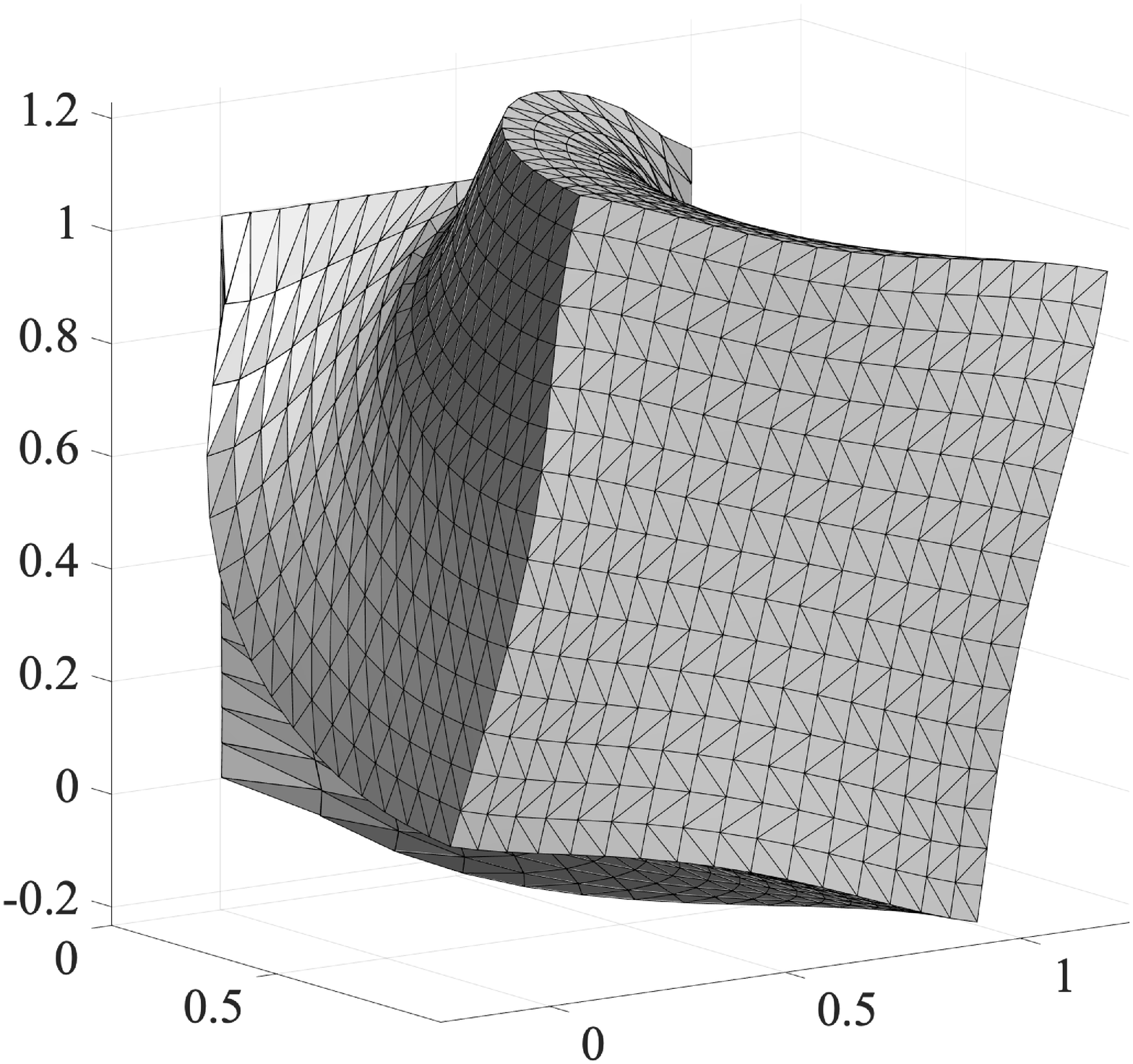}
	\end{center}
\caption{Twisting solution using continuous $P^1$ (left) and discontinuous $P^1$ (right) elements.\label{twistsol}}
\end{figure}

For our twisting example we use a domain $\Omega = (0,1)\times(0,1)\times(0,1)$ with $\bi u = \bi 0$ at $X = 0$ and a twisting volume force 
\[
\hat{\bi f} = (0,200-400 Z,400 Y-200) \quad \text{GN/m$^3$}
\]
The mesh is shown in Fig. \ref{fig:twistmesh}, and the solutions for continuous Galerkin (CG) and for DG are shown in Fig. \ref{twistsol}.
\subsubsection{Stretching}
\begin{figure}
	\begin{center}
	\includegraphics[scale=0.3]{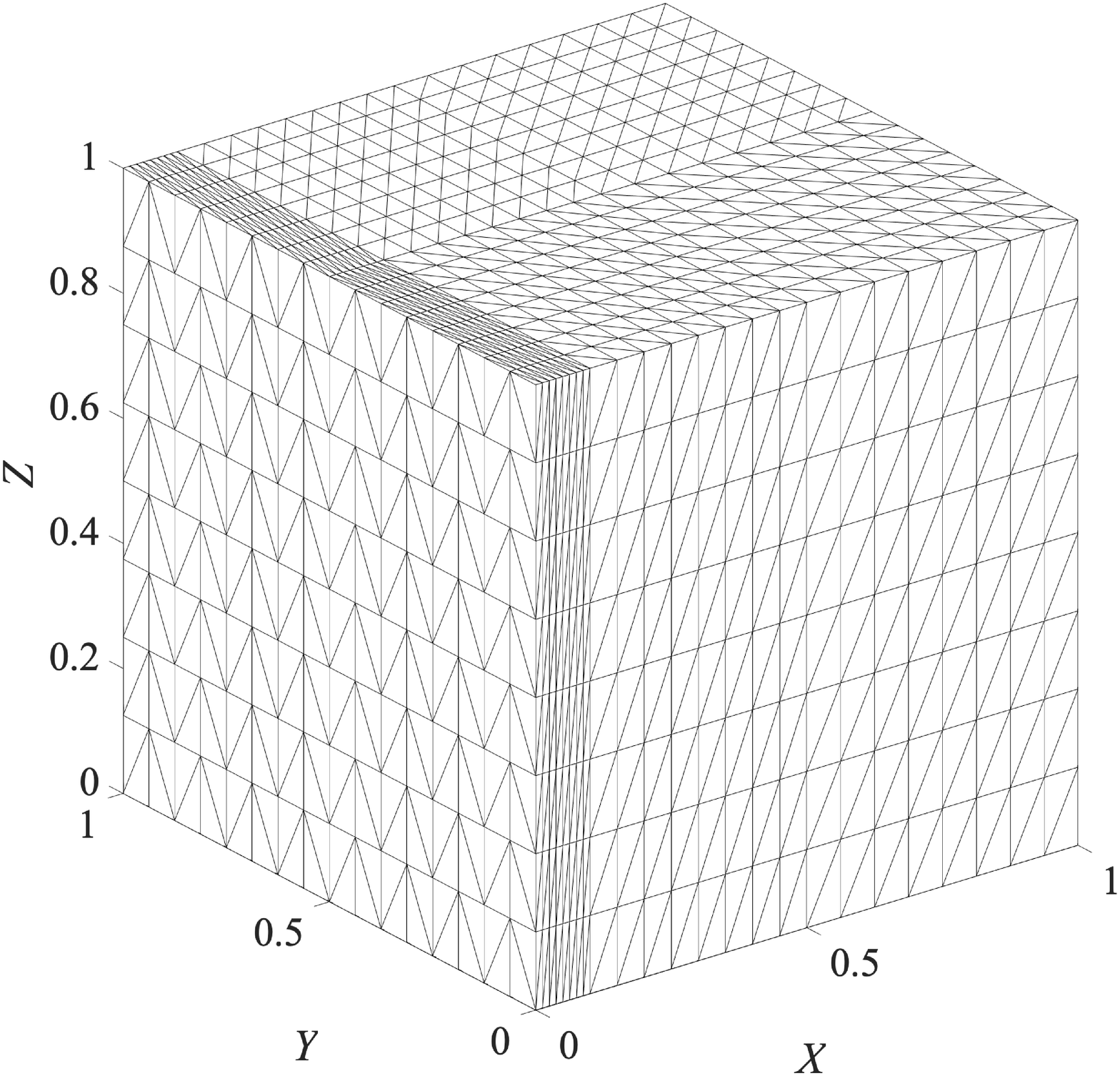}
	\end{center}
\caption{Mesh used for stretching computations.\label{fig:stretchmesh}}
\end{figure}

\begin{figure}
	\begin{center}
	\includegraphics[scale=0.2]{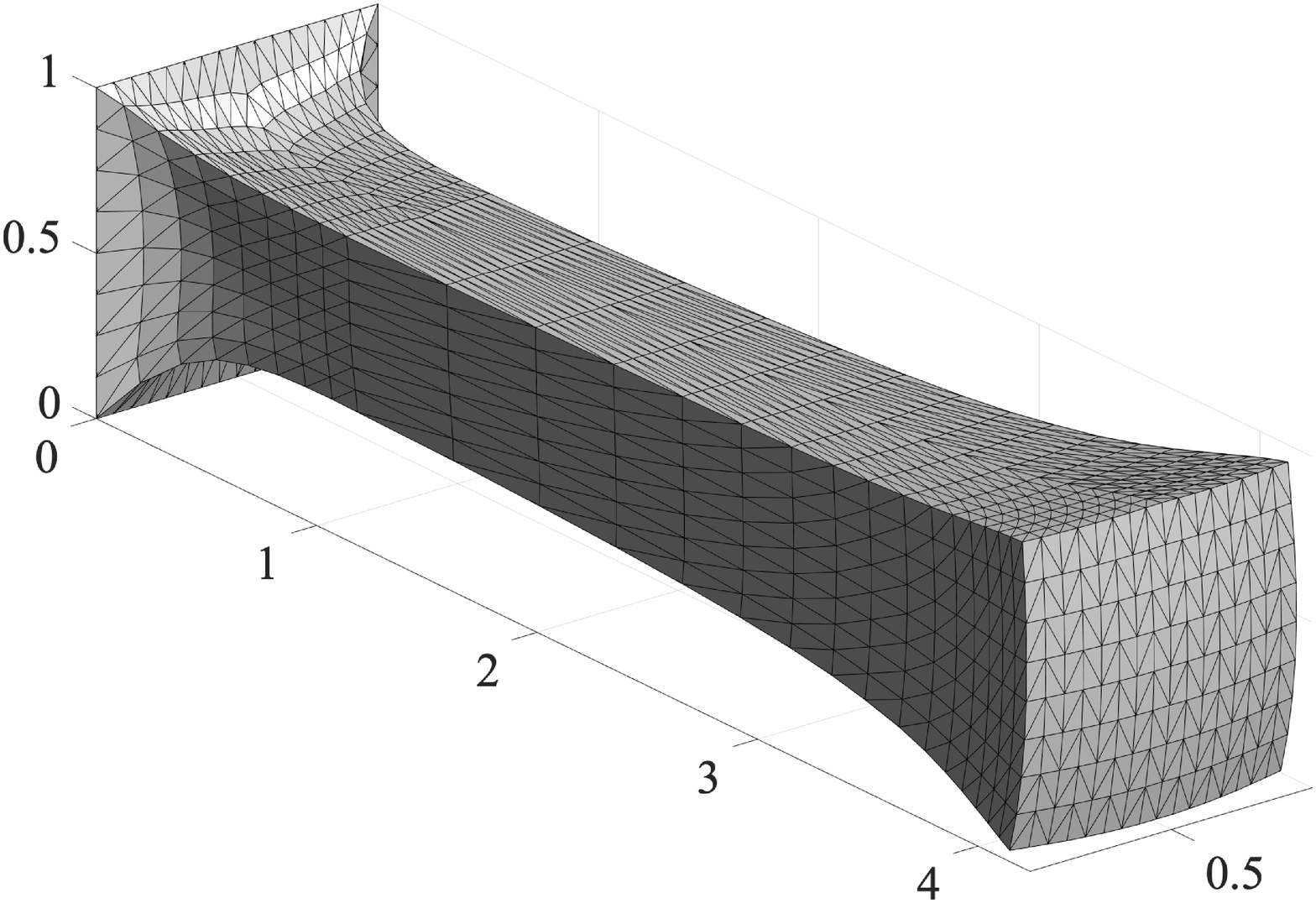}\includegraphics[scale=0.2]{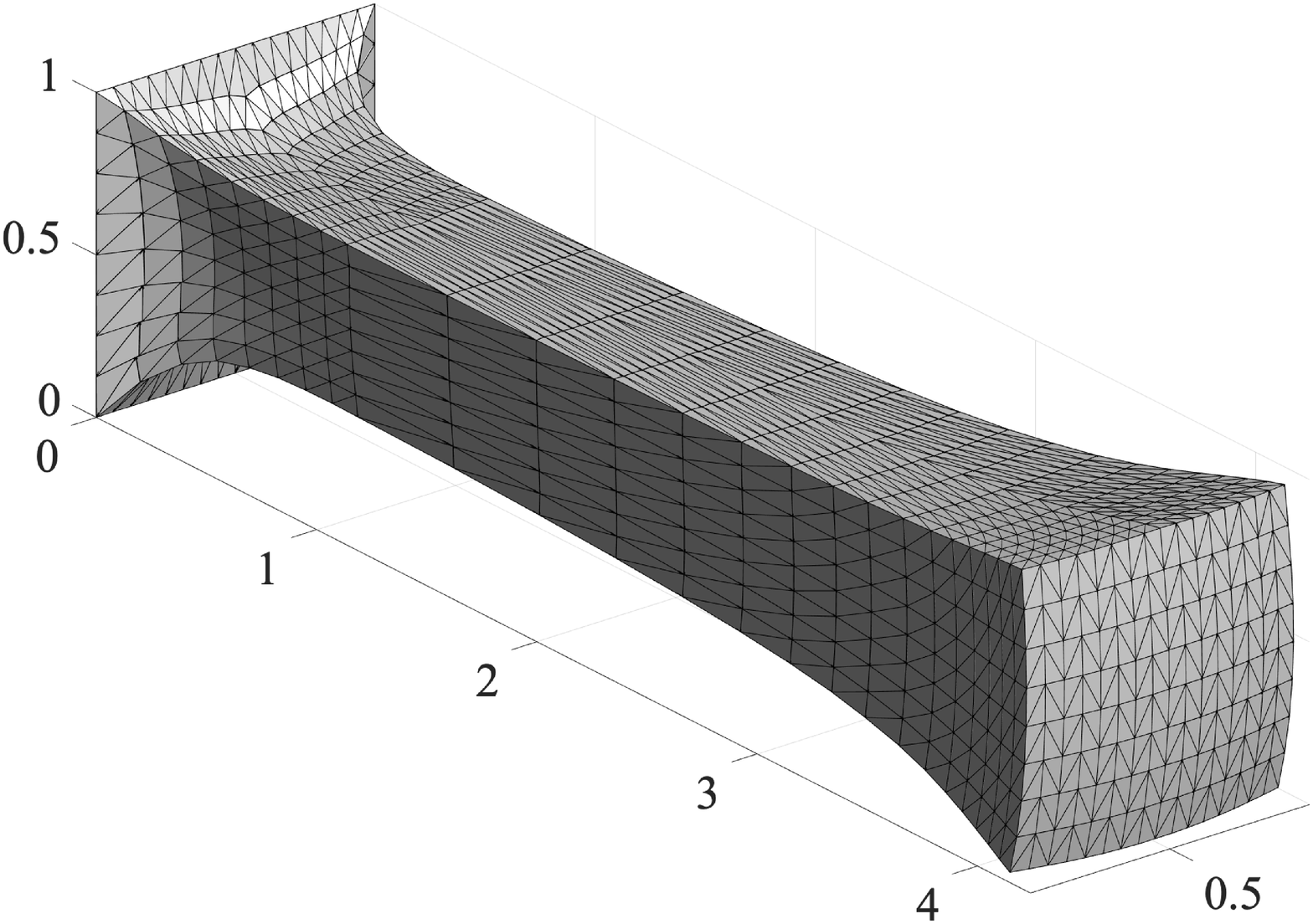}
	\end{center}
\caption{Stretching solution using continuous $P^1$ (left) and discontinuous $P^1$ (right) elements.\label{stretchsol}}
\end{figure}

The domain and boundary conditions are the same as in the twisting example but with a stretching load $\hat{\bi f} = (250,0,0)$ GN/m$^3$.
The mesh is shown in Fig. \ref{fig:stretchmesh}, and the solutions for CG and for DG are shown in Fig. \ref{stretchsol}.
\subsubsection{Bending}
\begin{figure}
	\begin{center}
	\includegraphics[scale=0.3]{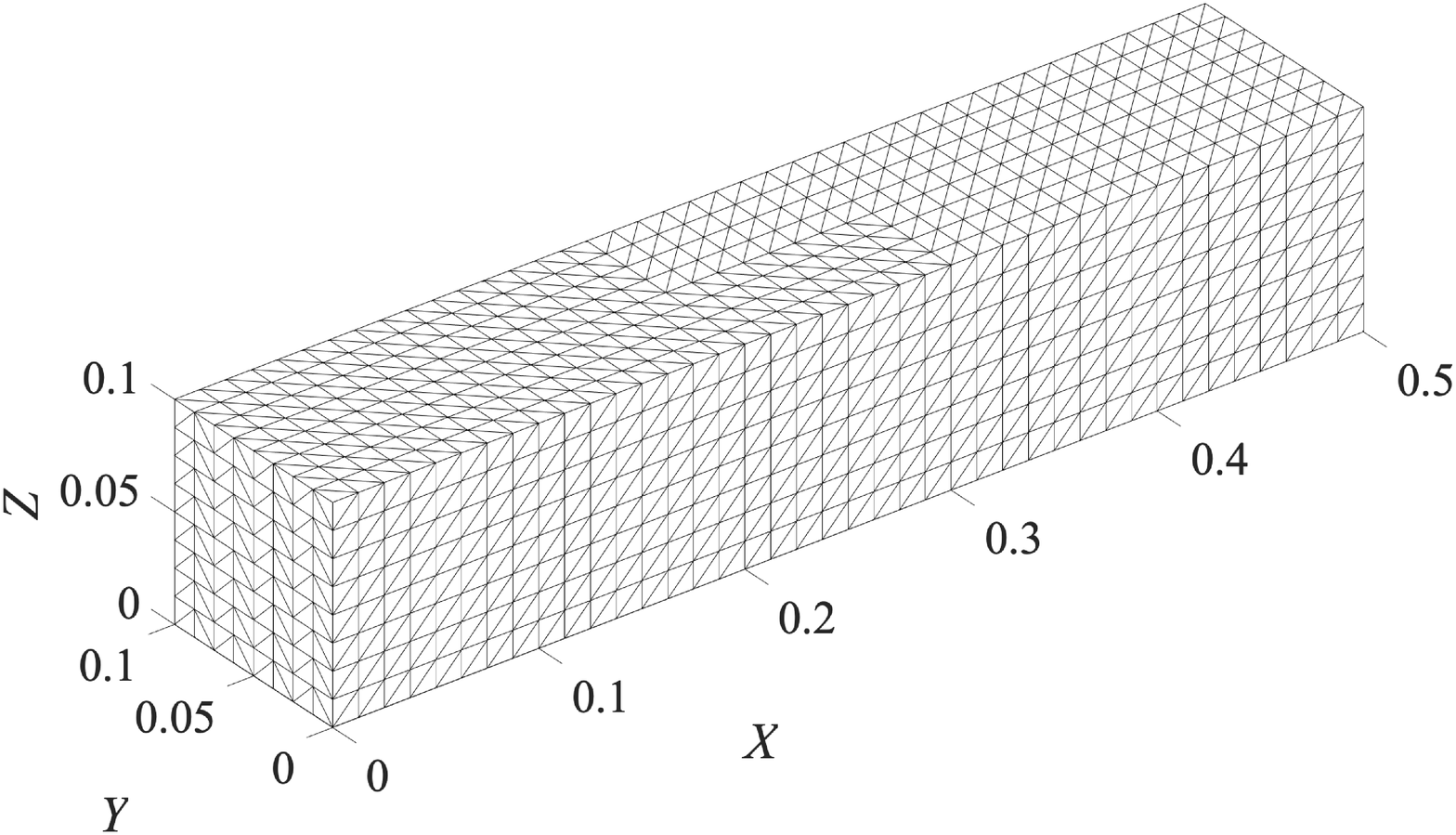}
	\end{center}
\caption{Mesh used for bending computations.\label{fig:bendmesh}}
\end{figure}

\begin{figure}
	\begin{center}
	\includegraphics[scale=0.2]{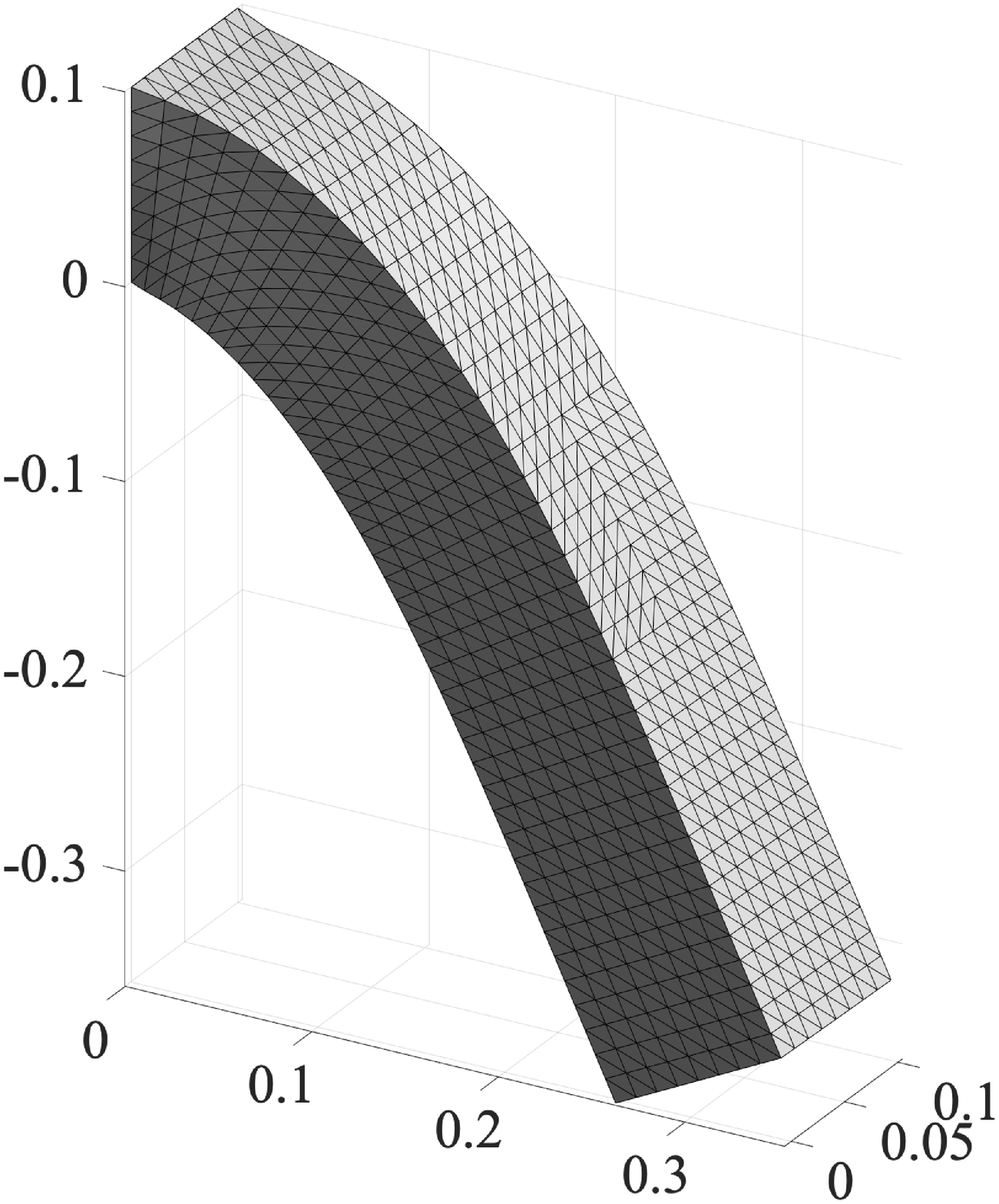}\includegraphics[scale=0.2]{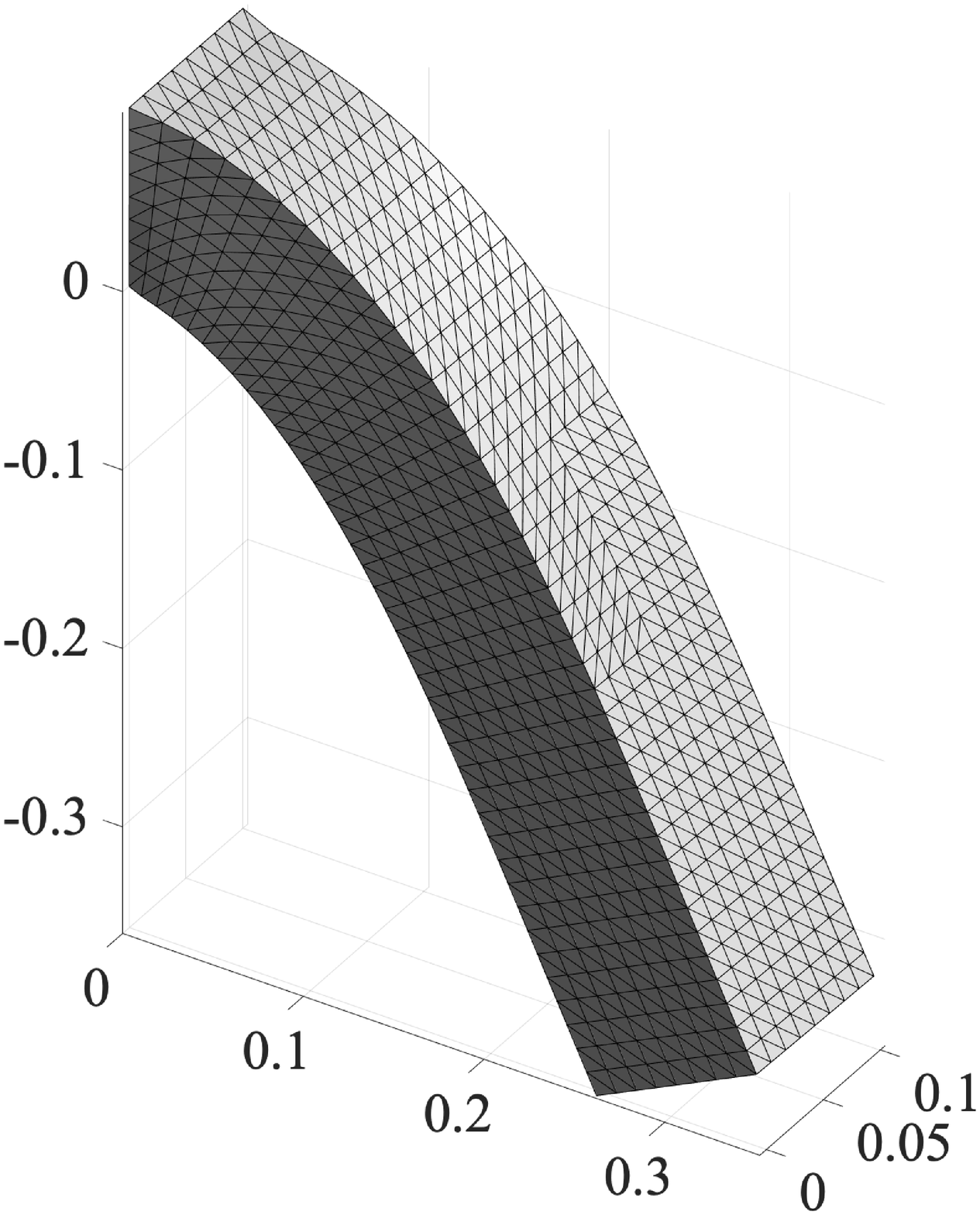}
	\end{center}
\caption{Bending solution using continuous $P^1$ (left) and discontinuous $P^1$ (right) elements.\label{bendsol}}
\end{figure}

The domain is $(0,1/2)\times(0,1/10)\times(1/10)$  with $\bi u = \bi 0$ at $X = 0$ and a bending load $\hat{\bi f} = (0,0,-15)$ GN/m$^3$.
The mesh is shown in Fig. \ref{fig:bendmesh}, and the solutions for CG and for DG are shown in Fig. \ref{bendsol}.

\paragraph{Acknowledgements}
This research was partly supported by the Swedish Research Council Grants Nos. 2017-03911, 2018-05262, 2021-04925, and the Swedish Research Programme Essence.

\newpage

\end{document}